\documentclass[12pt]{iopart}
\usepackage{pdflscape}
\usepackage{epsfig}
\usepackage{iopams}
\usepackage{cite}

\renewcommand{\theequation}{\thesection.\arabic{equation}}
\newcommand{\be}{\begin{equation}}
\newcommand{\ee}{\end{equation}}
\newcommand{\ben}{\begin{equation*}}
\newcommand{\een}{\end{equation*}}
\newcommand{\ba}{\begin{array}}
\newcommand{\ea}{\end{array}}
\newcommand{\g}{\mathfrak{g}}

\newtheorem{theorem}{Theorem}
\newtheorem{proposition}{Proposition}

\renewcommand{\theequation}{\thesection.\arabic{equation}}

\begin{document}

\title[On immersion formulas for soliton surfaces]
{On immersion formulas for soliton surfaces}
\author{A. M. Grundland$^{1,2}$, D. Levi$^3$ and L. Martina$^4$}

\address{$^1$ Centre de Recherches Math\'ematiques, Universit\'e de Montr\'eal,\\ Montr\'eal CP 6128 Succ. Centre-Ville (QC) H3C 3J7, Canada}
\address{$^2$ Department of Mathematics and Computer Science, Universit\'e du Qu\'ebec, Trois-Rivi\`eres, CP 500 (QC) G9A 5H7, Canada}
\address{$^3$ Dipartimento di Mathematica e Fisica dell'Universit\`a Roma Tre, Sezione INFN di Roma Tre, Via della Vasca Navale 84, Roma, 00146 Italy}
\address{$^4$ Dipartimento di Mathematica e Fisica dell'Universit\`a del Salento, Sezione INFN di Lecce, Via Arnesano, C.P. 193 Lecce, 73100 Italy}

\ead{grundlan@crm.umontreal.ca, decio.levi@roma3.infn.it, luigi.martina@le.infn.it}

\begin{abstract}
This paper is devoted to a study of the connections between three different analytic descriptions for the immersion functions of 2D-surfaces corresponding to the following three types of symmetries:  gauge symmetries of the linear spectral problem, conformal transformations in the spectral parameter and generalized symmetries of the associated integrable system. After a brief exposition of the theory of soliton surfaces and of the main tool used to study classical and generalized Lie symmetries, we derive the necessary and sufficient conditions under which the immersion formulas associated with these symmetries are linked by gauge transformations. We illustrate the theoretical results by examples involving the sigma model.

{\it This work is dedicated to Jiri Patera and Pavel Winternitz on the occasion of their 80th birthday.}

\paragraph{}Keyword: Integrable systems, Soliton surfaces, Immersion formulas, Generalized symmetries
\end{abstract}

\maketitle

\section{Introduction}\setcounter{equation}{0}
Soliton surfaces associated with integrable systems have been shown to play an essential role in many problems with physical applications (see e.g. \cite{Sym82,Sym85,Tafel95,Bobenko,Cieslinski97,Cieslinski07,DS92,FG96,FGFL,Grundland15,GPR14,GP11,GP12,Olver93,RS00,Helein01}). We say that a surface is integrable if the Gauss-Mainardi-Codazzi equations corresponding to it are integrable, i.e.  if  they can be represented as the compatibility conditions for some ``non-fake'' Linear Spectral Problem (LSP) \cite{Sym82,Sym85,Tafel95,Bobenko,Cieslinski97,cn91,hb, lsz90, lll10, m02, m10, m04, s01, s02,glsgal}. The possibility of using an LSP to represent a moving frame on the integrable surface has yielded many new results concerning the intrinsic geometric properties of such surfaces (see e.g. \cite{Olver93,Cartan53}). In the present state of development, it has proved most fruitful to extend such characterizations of soliton surfaces via their immersion functions (see e.g. \cite{BBT,Bobenko,Helein01,MS04,Mikhailov86,RS00} and references therein). 

The construction of surfaces related to completely integrable models was initated by A. Sym and J. Tafel \cite{Sym82,Sym85,Tafel95}. This construction makes use of the conformal invariance of the zero-curvature representation of the system with respect to the spectral parameter. Another approach for finding such surfaces has been formulated by J Cieslinski and A Doliwa \cite{Cieslinski97,Cieslinski07,DS92} using gauge symmetries of the LSP.
A third approach, using  the LSP for integrable systems and their symmetries has been introduced by Fokas and Gel'fand \cite{FG96,FGFL}, to construct families of soliton surfaces. Most recently, in a series of papers \cite{Grundland15,GPR14,GP11,GP12}, a reformulation and extension of the Fokas-Gel'fand immersion formula has been performed through the formalism of generalized vector fields and their actions on jet spaces. This extension has provided the necessary and sufficient conditions for the existence  of soliton surfaces in terms of the symmetries of the LSP and integrable models. 
The objective of this paper is to investigate and construct the relation between the three approaches concerning 2D-soliton surfaces associated with integrable systems. 

This paper is organized as follows. Section 2  contains a brief summary of the results concerning the construction of soliton surfaces and symmetries. 
Using the Sym-Tafel (ST) formula, we get the surface by differentiating the solution of the LSP with respect to the spectral parameter. Through the Cieslinski-Doliwa (CD) formula we apply a gauge transformation and through the Fokas-Gel'fand (FG) formula we consider the generalized symmetries of the associated integrable equation. 
In Section 3 we  demonstrate that the immersion problem can be mapped through a gauge to any of the three immersion formulas listed above and we show that these formulas correspond to possibly different parametrizations of the same surface. Then, in Section 4, we apply the results on the example of the sigma model. Section 5 contains the concluding remarks.

\section{Summary of results on the construction of soliton surfaces}\setcounter{equation}{0}
In this section we recall the main tools used to study symmetries suitable for the use of Fokas-Gel'fand formulas for the construction of 2D surfaces. We make use of the formalism of vector fields and their prolongations as presented in \cite{Olver93}. More specifically, we rewrite the formula for the immersion functions of 2D surfaces  in terms of the prolongation formalism of the vector fields instead of the  Fr\'echet derivatives.

\subsection{Classical and generalized Lie symmetries}\setcounter{equation}{0}
Let $X$ (with coordinates $x_\alpha$, $\alpha=1,...,p$) and $\mathcal{U}$ (with coordinates $u^k$, $k=1,...,q$) be differential manifolds representing spaces of independent and dependent variables, respectively. Let $J^n=J^n(X\times\mathcal{U})$ denote the $n$-jet space over $X\times\mathcal{U}$. The coordinates of $J^n$ are given by $x_\alpha$, $u^k$ and $u^k_J=\frac{\partial^nu^k}{\partial x_{j_1}...\partial x_{j_n}}$  where $J=(j_1,...,j_n)$ is a symmetric multi-index.  
We denote these coordinates by $x$ and $u^{(n)}$. 
On $J^n$ we can define  a system of partial differential equations (PDEs) in $p$ independent and $q$ dependent variables given by $m$ equations of the form 
\be
\Omega^\mu(x,u^{(n)})=0.\qquad \mu=1,...,m\label{1.1}
\ee

We consider a vector field $v$ tangent to $J^0=X\times\mathcal{U}$
\be
v=\xi^\alpha(x,u)\partial_{\alpha}+\varphi^k(x,u)\partial_{k},
\ee
where $\partial_{\alpha}={\partial}/{\partial x_\alpha}$, $\partial_{k}={\partial}/{\partial u^k}$ and we adopt the summation convention over repeated indices. Such a field defines vector fields, $\mbox{pr}^{(n)}v$ on $J^n$ \cite{Olver93}
\be
\mbox{pr}^{(n)}v=\xi^\alpha\partial_{\alpha}+\varphi_J^k\frac{\partial}{\partial u^k_J}.
\ee
The functions $\varphi^k_J$ are given by
\be
\varphi^k_J=D_JR^k+\xi^\alpha u^k_{J,\alpha}, \quad R^k=\varphi^k-\xi^\alpha u^k_\alpha,
\ee
where the operators $D_J$ correspond to multiple total derivatives, each of which is a combination of total derivatives of the form
\be
D_\alpha=\partial_{\alpha}+u^k_{J,\alpha}\frac{\partial}{\partial u^k_J},\hspace{8mm}\alpha=1,\ldots,p
\ee
 and $R^k$ are the so-called characteristics of the vector field $v$. In the following, the representation of $v$ can be written equivalently as
\be
v=\xi^\alpha D_\alpha+\omega_R,\qquad \omega_R=R^k\frac{\partial}{\partial u^k}.
\ee
One says that the vector field $v$ is a classical Lie point symmetry of a nondegenerate system of PDEs (\ref{1.1}) if and only if its $n$-th prolongation of $v$ is such that
\be
\mbox{pr}^{(n)}v\Omega^\mu(x,u^{(n)})=0, \qquad \mu=1,...,m
\ee
whenever $\Omega^\mu(x,u^{(n)})=0, \, \mu=1,...,m$ are satisfied. 
By a totally non-degenerate system we mean a system for which all their prolongations have maximal rank and are locally solvable \cite{Olver93}.
 It follows from the well-known properties of the symmetries of a differential system that the commutator of two  symmetries is again a symmetry. Thus,  such symmetries form a Lie algebra $\g$, which locally defines an action of a Lie group $G$ on $J^0$.

Every solution of (\ref{1.1}) can be represented by its graph, $u^k=\theta^k(x)$, which is a section of $J^0$. The symmetry group $G$ transforms solutions into solutions. This means that a graph corresponding to one solution is transformed into a graph associated with another solution. If the graph is preserved by the group $G$ or equivalently, if the vector fields $v$ from the algebra $\g$ are tangent to the graph, then the related solution is said to be $G$-invariant. Invariant solutions satisfy, in addition to the equations (\ref{1.1}), the characteristic equations equated to zero
\be
\varphi^k_a(x,\theta)-\xi^\alpha_a(x,\theta)\theta^k_{,\alpha}=0,\qquad a=1,...,r
\ee
where the index $a$ runs over the  generators of $\g$.

It may happen that invariant solutions are restricted in number or trivial if the full symmetry group is small. To extend the number of symmetries, and thus of solutions, one looks for generalized symmetries. They exist only if the nonlinear equation (\ref{1.1}) is integrable \cite{MSS}, i.e. it has been obtained as the compatibility of a Lax pair (see (\ref{2.3}) and in the following). A generalized vector field is expressed in terms of the characteristics
\be
\omega_R=R^k[u]\frac{\partial}{\partial u^k}
,\label{1.12}
\ee
where $[u]=(x,u^{(n)})\in J^n.$
The prolongation of an evolutionary vector field $\omega_R$ is given by
\be
\mbox{pr}\omega_R=\omega_{R}+D_JR^k\frac{\partial}{\partial u_J^k}.
\ee
A vector field $\omega_R$ is a generalized symmetry of a nondegenerated system of PDEs (\ref{1.1}) if and only if \cite{Olver93}
\be
\mbox{pr}\omega_R\Omega^\mu(x,u^{(n)})=0,\label{2.10*}
\ee
whenever $\Omega(x,u^{(n)})=0$ and its differential consequences are satisfied.

\subsection{The immersion formulas for soliton surfaces}
In order to  analyse  the Fokas-Gel'fand immersion formula for a surface in $2D$, we briefly summarize the results obtained in \cite{Sym82,Sym85,Tafel95,Bobenko,Cieslinski97,Cieslinski07,DS92,FG96,FGFL,Grundland15,GPR14,GP11,GP12}.

Let us consider an integrable system of partial differential equations (PDEs) in two independent variables $x_1,x_2$ and $m$ dependent variables $u^k(x_1,x_2)$ written as
\be
\Omega[u]=0.\label{2.1}
\ee
Suppose that the system (\ref{2.1}) is obtained as the compatibility of a matrix LSP written in the form \cite{Sym82}
\be
\partial_\alpha\Phi(x_1,x_2,\lambda)-U_\alpha([u],\lambda)\Phi(x_1,x_2,\lambda)=0,\qquad\alpha=1,2\label{2.3}
\ee
In what follows, the potential matrices $U_\alpha$ can be defined on the extended jet space $\mathcal{N}=(J^n,\lambda)$, where $\lambda$ is the spectral parameter. The compatibility condition of the LSP (\ref{2.3}), often called the Zero-Curvature Condition (ZCC)
\be
D_2U_1-D_1U_2+[U_1,U_2]=0, \label{2.4}
\ee
which is assumed to be valid for all values of $\lambda$, implies (\ref{2.1}). The bracket in (\ref{2.4}) denotes the Lie algebra commutator. Equation (\ref{2.4}) provides a representation for the initial system (\ref{2.1}) under consideration. The $m$-dimensional matrix functions $U_\alpha$ take values in some semisimple Lie algebra $\mathfrak{g}$ and the wavefunction $\Phi$ takes values in the corresponding Lie group $G$. 
We can say that, as long as the potential matrices $U_\alpha([u],\lambda)$ satisfy the ZCC (\ref{2.4}), there exists a group-valued function $\Phi$ which satisfies (\ref{2.3}). There exists a subclass of $\Phi$ which can be defined formally on the extended jet space $\mathcal{N}$. For $\Phi$ belonging to this subclass we can write formally $\Phi=\Phi([u],\lambda)\in G$, meaning that $\Phi$ depends functionally on $[u]$ and meromorphically in $\lambda$.
When $\Phi=\Phi([u],\lambda)$ the LSP (\ref{2.3}) can be written as
\be
\Lambda_\alpha([u],\lambda)\equiv D_\alpha\Phi([u],\lambda)-U_\alpha([u],\lambda)\Phi([u],\lambda)=0,\qquad\alpha=1,2\label{2.6}
\ee
which is a convenient form for the analysis we carry out in the following. 

In reference \cite{FGFL}, the authors looked for a simultaneous infinitesimal deformation of the associated LSP (\ref{2.6}) which preserves the ZCC (\ref{2.4})  
\be
\left(\ba{c}
\tilde{U}_1\\
\tilde{U}_2\\
\tilde{\Phi}
\ea\right)=\left(\ba{c}
U_1\\
U_2\\
\Phi
\ea\right)+\epsilon\left(\ba{c}
A_1\\
A_2\\
\Psi
\ea\right)+O(\epsilon^2),\label{2.7}
\ee
where the matrices $\tilde{U}_1$, $\tilde{U}_2$, $A_1$ and $A_2$ take values in the Lie algebra $\g$, while $\tilde{\Phi}$ and $\Psi$ belong to the corresponding Lie group $G$. The parameter $\lambda$ is left invariant under the transformation (\ref{2.7}) and $0<\epsilon\ll1$. The corresponding infinitesimal generator formally takes the evolutionary form
\be
\hat{X}_e=A^1\partial_{U_1}+A^2\partial_{U_2}+\Psi\partial_\Phi.\label{mystery}
\ee
Equation (\ref{mystery}) can be written as 
\be
\hat{X}_\epsilon=A^{1j}\partial_{U_1^j}+A^{2j}\partial_{U_2^j}+\Psi^j\partial_{\Phi^j},\label{2.8}
\ee
where we decompose the matrix functions $A^1$ and $A^2$ in the basis $e_j$, $j=1,...,s$ for the Lie algebra $\mathfrak{g}$
\begin{equation}
U_\alpha=U_\alpha^je_j\in\mathfrak{g},\qquad [e_i,e_j]=c_{ij}^ke_k
\end{equation}
where [ , ] is the Lie algebra commutator and $c_{ij}^k$ are the structural constants of $\mathfrak{g}$. 
Since this generator $\hat{X}_e$ does not transform $\lambda$, these symmetries preserve the singularity structure of the potential matrices $U_\alpha$ in the spectral parameter $\lambda$. 
The infinitesimal deformation of the LSP (\ref{2.3}) and the ZCC (\ref{2.4}) under the infinitesimal transformation (\ref{2.7}) requires that the matrix functions $U_\alpha$ and $\Psi$ satisfy, at first order in $\epsilon$, the equations
\be
D_\alpha\Psi=U_\alpha\Psi+A_\alpha\Phi,\qquad \alpha=1,2\label{2.9i}
\ee
and
\be
D_2A_1-D_1A_2+[A_1,U_2]+[U_1,A_2]=0,\label{2.9ii}
\ee
The equation (\ref{2.9ii}) coincides with the compatibility condition for (\ref{2.9i}). 
For the given matrix functions $U_{\alpha}$, $A_{\alpha}\in\g$ and $\Phi\in G$ satisfying equations (\ref{2.4}), (\ref{2.6}) and (\ref{2.9ii}), an infinitesimal symmetry of the matrix system of the integrable PDEs (\ref{2.1}) allows us to generate a 2D-surface immersed in the Lie algebra $\g$. According to \cite{FG96} this result is formulated as follows. 

\begin{theorem} 
If the matrix functions $U_\alpha\in\g$, $\alpha=1,2$ and $\Phi\in G$ of the LSP (\ref{2.6}) satisfy the ZCC (\ref{2.4})  and $A_\alpha\in\g$ are linearly independent matrix functions which satisfy (\ref{2.9ii}) and $\Phi\in G$ satisfies the LSP (\ref{2.6}), then there exists (up to affine transformations) a 2D-surface with a $\g$-valued immersion function $F([u],\lambda)$ such that the tangent vectors to this surface are given by
\be
D_\alpha F([u],\lambda)=\Phi^{-1}A_\alpha([u],\lambda)\Phi,\qquad\alpha=1,2\label{2.11}
\ee
\end{theorem}

{\bf Proof} The compatibility condition of (\ref{2.11}) coincides with (\ref{2.9ii}). So an immersion function $F([u],\lambda)$ exists and can be assumed to take its values in the Lie algebra $\g$. If we define the matrix function
\be
\Psi=\Phi F,\label{psi}
\ee
then, using (\ref{2.11}), the function $\Psi$ satisfies (\ref{2.9i}). Hence, since $F=\Phi^{-1}\Psi$, the formula $\tilde{\Phi}=\Phi+\varepsilon\Psi=\Phi(\mathbb{I}+\varepsilon F)$ implies that $\tilde{\Phi}$ is in the Lie group $G$.
The immersion function
\begin{equation}
F=\left.\left(\Phi^{-1}([u],\lambda)\cdot\frac{d\tilde{\Phi}([u],\lambda,\epsilon)}{d\epsilon}\right)\right\vert_{\epsilon=0}
\end{equation}
is an element of the Lie algebra $\mathfrak{g}$.
This shows that we have constructed an appropriate infinitesimal deformation of the wavefunction $\Phi$.\hfill$\square$

In \cite{Sym82,Cieslinski97} it was shown  that the admissible symmetries of the ZCC (\ref{2.4}) include a conformal transformation of the spectral parameter $\lambda$, a gauge transformation of the wavefunction $\Phi$ in the LSP (\ref{2.3}) and generalized symmetries of the integrable system (\ref{2.1}). All these symmetries can be used to determine explicitly a $\g$-valued immersion function $F$ of a 2D-surface. Thus, a  generalization of the FG formula for immersion can be formulated as follows \cite{FGFL,Grundland15}.

\begin{theorem}
Let the set of scalar functions $\lbrace u^k\rbrace$ satisfy a system of integrable PDEs $\Omega[u]=0$. Let the $G$-valued function $\Phi([u],\lambda)$ satisfy the LSP (\ref{2.6}) of $\g$-valued potentials $U_\alpha([u],\lambda)$. Let us define the linearly independent $\g$-valued matrix functions $A_\alpha([u],\lambda)$ $(\alpha=1,2)$  by the equations
\be
\hspace{-2.5cm}A_\alpha([u],\lambda)=\beta(\lambda)D_\lambda U_\alpha+\left(D_\alpha S+[S,U_\alpha]\right) +\mbox{pr}\omega_RU_\alpha+\left(\mbox{pr}\omega_R(D_\alpha\Phi-U_\alpha\Phi)\right)\Phi^{-1}.\label{2.12}
\ee
Here $\beta(\lambda)$ is an arbitrary scalar function of $\lambda$, $S=S([u],\lambda)$ is an arbitrary $\g$-valued matrix function defined on the jet space  $\mathcal{N}$, $\omega_R=R^k[u]\partial_{u^k}$ is the  vector field, written in evolutionary form, 
of the generalized symmetries of the integrable PDEs $\Omega[u]=0$ given by the ZCC (\ref{2.4}). Then there exists a 2D-surface with immersion function $F([u],\lambda)$ in the Lie algebra $\g$ given by the formula (up to an additive $\g$-valued constant)
\be
F([u],\lambda)=\Phi^{-1}\left(\beta(\lambda)D_\lambda\Phi+S\Phi+\mbox{pr}\omega_R\Phi\right).\label{2.14}
\ee
The integrated form of the surface (\ref{2.14}) defines a mapping $F:\mathcal{N}\rightarrow\g$ and we will refer to it as the ST immersion formula (when $S=0,\omega_R=0$) \cite{Sym82,Sym85,Tafel95}
\be
F^{ST}([u],\lambda)=\beta(\lambda)\Phi^{-1}(D_{\lambda}\Phi)\in\g\label{fst},
\ee
the CD immersion formula (when $\beta=\omega_R=0$) \cite{Cieslinski97,Cieslinski07,DS92}
\be
F^{CD}([u],\lambda)=\Phi^{-1}S([u],\lambda)\Phi\in\g\label{fcd},
\ee
or the FG immersion formula (when $\beta=0,S=0$) \cite{FG96,FGFL}
\be
F^{FG}([u],\lambda)=\Phi^{-1}(\mbox{pr}\omega_R\Phi)\in\g\label{ffg}.
\ee
\end{theorem}

\subsection{Application of the method}
The  construction of soliton surfaces requires three elements for an explicit representation of the immersion function $F\in\g$: 
\begin{enumerate}
\item An LSP (\ref{2.3}) for the integrable PDE.\label{item1}
\item A generalized symmetry $\omega_R$ of the integrable PDE.
\item A solution $\Phi$ of the  LSP associated with the soliton solution of the integrable PDE.
\end{enumerate}

Note that item \ref{item1} is always required. In its presence, even without one of the remaining two objects, we can obtain an immersion function $F$.

When a solution $\Phi$ of the LSP is unknown,  the geometry of the surface $F$ can be obtained using the non-degenerate Killing form on the Lie algebra $\g$. The 2D-surface with the immersion function $F$ can be interpreted as a pseudo-Riemannian manifold. 

When the generalized symmetries $\omega_R$ of the integrable PDE are unknown but we know a solution $\Phi$ of the LSP then we can define the 2D-soliton surface  using the gauge transformation and the $\lambda$-invariance of the ZCC
\be \label{terza}
F=\Phi^{-1}(\beta(\lambda)D_\lambda\Phi+S\Phi),
\ee
where $\beta(\lambda)$ is an arbitrary scalar function of $\lambda$ and $S$ is an arbitrary $\g$-valued matrix function defined on the extended jet space $\mathcal{N}$. Equation (\ref{terza}) is consistent with the tangent vectors
\be
D_\alpha F=\beta(\lambda)\Phi^{-1}(D_\lambda U_\alpha)+\Phi^{-1}(D_\alpha S+[S,U_\alpha])\Phi.
\ee

In all cases, the tangent vectors, given by (\ref{2.11}), and the unit normal vector to a 2D-surface expressed in terms of matrices are
\be \label{primo}
D_\alpha F=\Phi^{-1}A_\alpha\Phi\in\g,\qquad N=\frac{\Phi^{-1}[A_1,A_2]\Phi}{(\frac{1}{2}\mbox{tr}[A_1,A_2]^2)^{1/2}}\in\g,
\ee
where the $\g$-valued matrices $A_{\alpha}$ are given by (\ref{2.12}).
The first and second fundamental forms are given by
\begin{equation}
I=g_{ij}dx_idx_j,\qquad II=b_{ij}dx_idx_j,\qquad i=1,2\label{I}
\end{equation}
where
\begin{equation}
\hspace{-1cm}g_{ij}=\frac{\epsilon}{2}\tr(A_iA_j)),\qquad b_{ij}=\frac{\epsilon}{2}\tr((D_jA_i+[A_i,U_j])N),\qquad\epsilon=\pm1\label{II}
\end{equation}
This gives the following expressions for the mean and Gaussian curvatures
\begin{equation*}
\hspace{-2.5cm}\begin{array}{l}
H=\frac{1}{\Delta}\left\lbrace\tr(A_2^2)\tr((D_1A_1+[A_1,U_1])N)-8\tr(A_1A_2)\tr((D_2A_1+[A_1,U_2])N)\right.\\
\hspace{1cm}\left.+\tr(A_1^2)\tr((D_2A_2+[A_2,U_2])N)\right\rbrace,\\
K=\frac{1}{\Delta}\left\lbrace\tr((D_1A_1+[A_1,U_1])N)\tr((D_2A_2+[A_2,U_2])N)-2\tr^2((D_2A_1+[A_1,U_2])N)\right\rbrace,\\
\end{array}
\end{equation*}

\begin{equation}
\hspace{-2.3cm}\Delta=\tr(A_1^2)\tr(A_2^2)-4\tr(A_1A_2),
\label{KH}
\end{equation}
which are expressible in terms of $U_\alpha$ and $A_\alpha$ only.

The study of soliton surfaces defined via the FG formula for immersion provides a unique mechanism for studying the relationship between these various characteristics of integrable systems. 
For example, do the infinite families of conservation laws have a geometric characterization? What is the geometry behind the Hamiltonian structure? Is there a geometric interpretation of the family of surfaces and frames associated with the spectral parameter? These answers will not only serve to construct surfaces with interesting geometric quantities but can also help to clarify some problems in the theory of integrable system.

Following the three terms in (\ref{2.14}) for the immersion of 2D-soliton surfaces in Lie algebras, we now show that there exists a relation between the Sym-Tafel, the Cieslinski-Doliwa and the Fokas-Gel'fand formulas.

\section{Mapping between the Sym-Tafel, the Cieslinski-Doliwa and the Fokas-Gel'fand immersion formulas.}\setcounter{equation}{0}
\subsection{$\lambda$-conformal symmetries and gauge transformations}
In this subsection we show that the ST immersion formula can always be represented by a gauge transformation through the CD formula for immersion. The converse statement is also true: from a specific gauge it is always possible to determine the ST immersion formula for soliton surfaces.

\begin{proposition}
A symmetry of the ZCC (\ref{2.4}) of the LSP associated with an integrable system $\Omega[u]=0$ is a $\lambda$-conformal symmetry if and only if there exists a $\mathfrak{g}$-valued matrix function $S_1=S_1([u],\lambda)$ which is a solution of the system of differential equations
\be
D_\alpha S_1+[S_1,U_\alpha]=\beta(\lambda)D_\lambda U_\alpha.\qquad \alpha=1,2\label{3.1}
\ee
\end{proposition}

\textbf{Proof.} 
First we show that for any $\lambda$-conformal symmetry of the ZCC of the LSP associated with the integrable system $\Omega[u]=0$, there exists a $\g$-valued matrix function $S_1=S_1([u],\lambda)$ which is a solution of the system of differential equations (\ref{3.1}). Indeed, the linearly independent $\g$-valued matrix functions
\be
A_\alpha([u],\lambda)=\beta(\lambda)D_\lambda U_\alpha([u],\lambda),\label{3.2}
\ee
associated with the $\lambda$-conformal symmetry of the ZCC (\ref{2.4}) satisfy the infinitesimal deformation of the ZCC (\ref{2.9ii}) and the corresponding ST immersion function is
\be
F^{ST}([u],\lambda)=\beta(\lambda)\Phi^{-1}D_\lambda\Phi\in\g,\label{3.3}
\ee
with linearly independent tangent vectors
\be
D_\alpha F^{ST}=\beta(\lambda)\Phi^{-1}(D_\lambda U_\alpha)\Phi,\qquad\alpha=1,2.\label{3.4}
\ee

On the other hand any $\g$-valued matrix function can be written as the adjoint group action on its Lie algebra. This implies the existence of a $\g$-valued matrix function $S_1([u],\lambda)$ for which the ST immersion formula (\ref{3.3}) is the CD formula, i.e.
\be
F^{CD}([u],\lambda)=\Phi^{-1}S_1([u],\lambda)\Phi\in\g,\label{3.5}
\ee
whose tangent vectors are found to be
\be
D_\alpha F^{CD}=\Phi^{-1}(D_\alpha S_1+[S_1,U_\alpha])\Phi,\qquad\alpha=1,2.\label{3.6}
\ee
By comparing the tangent vectors (\ref{3.4}) and (\ref{3.6}) we obtain (\ref{3.1}). It remains to show that the system (\ref{3.1}) is a solvable one. Indeed, from the compatibility condition of (\ref{3.1}) we get
\be\hspace{-2cm}\ba{l}
\beta(\lambda)D_2(D_\lambda U_1)-\beta(\lambda)D_1(D_\lambda U_2)-[\beta(\lambda)D_\lambda U_2-[S_1,U_2],U_1]-[S_1,D_2U_1]\\
+[\beta(\lambda)D_\lambda U_1-[S_1,U_1],U_2]+[S_1,D_1U_2]=0,\label{3.7}
\ea\ee
which has to be satisfied whenever (\ref{3.1}) holds. Using the ZCC (\ref{2.4}) and the Jacobi identity it is easy to show that (\ref{3.7}) is identically satisfied. So if we can find a gauge $S_1([u],\lambda)$ which satisfies (\ref{3.1}), then the ST immersion formula (\ref{3.3}) can always be represented by a gauge.

Conversely, we show that for any $\g$-valued matrix function $S_1$ defined as a solution of the system of PDEs (\ref{3.1}), there exists a $\lambda$-conformal symmetry of the ZCC of the LSP associated with the integrable system $\Omega[u]=0$. Indeed, comparing the immersion formulas (\ref{3.3}) with (\ref{3.5}) we find a linear matrix equation for the wavefunction $\Phi$
\be
\beta(\lambda)D_\lambda\Phi=S_1([u],\lambda)\Phi.\label{3.8}
\ee
If the gauge function $S_1([u],\lambda)$ is known, by solving (\ref{3.8}) we can  determine the wavefunction $\Phi$ and consequently obtain the ST immersion formula for 2D-soliton surfaces. Therefore, the ST formula for immersion (\ref{fst})
is equivalent to the CD immersion formula (\ref{fcd}) for the gauge $S_1$, which satisfies differential equation (\ref{3.1}). \hfill$\square$

\subsection{Generalized symmetries and gauge transformations}
In this subsection we discuss the links between gauge transformations and generalized symmetries of the ZCC associated with the integrable partial differential system $\Omega[u]=0$. We show that the immersion formula associated with the generalized symmetries (\ref{ffg}) can always be obtained by a gauge transformation and the converse statement is also true.

\begin{proposition}
A vector field $\omega_R$ is a generalized symmetry of the ZCC (\ref{2.4}) of the LSP associated with an integrable system $\Omega[u]=0$ if and only if there exists a $\mathfrak{g}$-valued matrix function (gauge) $S_2=S_2([u],\lambda)$ which is a solution of the system of differential equations.
\be
D_\alpha S_2+[S_2,U_\alpha]=\mbox{pr}\omega_RU_\alpha+\left(\mbox{pr}\omega_R(D_\alpha\Phi-U_\alpha\Phi)\right)\Phi^{-1}.\label{4.1}
\ee
\end{proposition}

\textbf{Proof.} First we demonstrate that for every infinitesimal generator $\omega_R$ which is a generalized symmetry of the ZCC (\ref{2.4}), there exists a $\g$-valued matrix function (gauge) $S_2=S_2([u],\lambda)$ which is a solution of the system of differential equations (\ref{4.1}). Indeed an evolutionary vector field $\omega_R$ is a generalized symmetry of the ZCC (\ref{2.4}) if and only if
\be
\mbox{pr}\omega_R(D_2U_1-D_1U_2+[U_1,U_2])=0,\label{4.2}
\ee
whenever the ZCC (\ref{2.4}) holds. Equation (\ref{4.2}) is equivalent to the infinitesimal deformation of the ZCC (\ref{2.4}) given by (\ref{2.9ii}), with linearly independent $\g$-valued matrix functions
\be
A_\alpha([u],\lambda)=\mbox{pr}\omega_RU_\alpha+(\mbox{pr}\omega_R(D_\alpha\Phi-U_\alpha\Phi))\Phi^{-1}.\qquad \alpha=1,2\label{4.4}
\ee
In the derivation of (\ref{4.4}) we use the fact that the total derivatives $D_\alpha$ commute with the prolongation of a vector field $\omega_R$ written in the evolutionary form \cite{Olver93}
\be
[D_\alpha,\mbox{pr}\omega_R]=0,\qquad\alpha=1,2.\label{4.5}
\ee
Using the LSP (\ref{2.6}), equation (\ref{4.4}) can be written in the equivalent form
\be
A_\alpha([u],\lambda)=\left[-U_\alpha(\mbox{pr}\omega_R\Phi)+\mbox{pr}\omega_R(D_\alpha\Phi)\right]\Phi^{-1}.\qquad\alpha=1,2 \label{4.6}
\ee
Substituting (\ref{4.6}) into (\ref{2.9ii}) we obtain
\ben
(-D_2U_1+D_1U_2-[U_1,U_2])(\mbox{pr}\omega_R\Phi)\Phi^{-1}=0,
\een
which is satisfied identically whenever the ZCC (\ref{2.4}) holds. 

An integrated form of the immersion function $F^{FG}([u],\lambda)$ of a 2D-surface associated with a generalized symmetry $\omega_R$ of the ZCC (\ref{2.4}) and the tangent vectors (\ref{2.11})
is given by the FG formula
\be
F^{FG}([u],\lambda)=\Phi^{-1}(\mbox{pr}\omega_R\Phi)\in\g.\label{4.8}
\ee
The fact that any $\g$-valued matrix function can be written under the adjoint group action implies that there exists a $\g$-valued gauge $S_2$, such that (\ref{3.5}) holds for the CD immersion function and its tangent vectors $D_\alpha F^{CD}$ given by (\ref{3.6}). Comparing equations (\ref{2.11}) and (\ref{4.4}) with (\ref{3.6}) we get (\ref{4.1}).

Let us show that the system (\ref{4.1}) always possesses a solution. The compatibility condition of (\ref{4.1}), whenever (\ref{4.1}) and (\ref{2.9ii}) hold, implies the relation
\be
[S_2,D_2U_1-D_1U_2]+[[S_2,U_1],U_2]-[[S_2,U_2],U_1]=0,\label{4.9}
\ee
which is identically satisfied in view of the ZCC (\ref{2.4}) and the Jacobi identity. So, if we can find a gauge function $S_2([u],\lambda)$ which satisfies (\ref{4.1}), then the FG formula (\ref{4.8}) can always be represented by a gauge transformation.

The converse statement is also true. We show that for any $\mathfrak{g}$-valued matrix function $S_2$ defined as a solution of the system of PDEs (\ref{4.1}), there exists a generalized symmetry $\omega_R$ of the ZCC of the LSP associated with $\Omega[u]=0$. Indeed, let the LSP of $\Omega[u]=0$ admit a gauge symmetry. If the gauge $S_2([u],\lambda)$ is given, then the immersion function $F^{CD}$ of a 2D-surface can be integrated explicitly \cite{Cieslinski97,Cieslinski07,DS92}
\begin{equation}
F^{CD}([u],\lambda)=\Phi^{-1}S_2([u],\lambda)\Phi\in\mathfrak{g},
\end{equation}
whenever the tangent vectors
\begin{equation}
D_\alpha F^{CD}=\Phi^{-1}(D_\alpha S_2+[S_2,U_\alpha])\Phi.
\end{equation}
are linearly independent.
It is straightforward to verify that the characteristics of a generalized vector field $\omega_R^S$, written in evalutionary form, associated with a gauge symmetry $S_2$, can be expressed as
\begin{equation}
A_\alpha=D_\alpha S_2+[S_2,U_\alpha]\in\mathfrak{g}.
\end{equation}
The matrices $A_\alpha$ identically satisfy the determining equations (\ref{2.9ii}) which are required for $\omega_R^S$ to be a generalized symmetry of the ZCC $\Omega[u]=0$
\begin{equation}
\hspace{-2cm}D_2A_1-D_1A_2+[A_1,U_2]+[U_1,A_2]=\mbox{pr}\omega_R^S(D_2U_1-D_1U_2+[U_1,U_2])=0,
\end{equation}
whenever $\Omega[u]=0$ holds. Hence, the vector field $\omega_R^S$ associated with a gauge symmetry $S_2$ is given by
\begin{equation}
\omega_R^S=(D_\alpha S_2+[S_2,U_\alpha])^j\frac{\partial}{\partial U_\alpha^j},
\end{equation}
where we have decomposed the matrix functions $A_\alpha$ and $U_\alpha$ in the basis $\lbrace e_j\rbrace_1^n$ for the Lie algebra
\begin{equation}\ba{l}
U_\alpha=U_\alpha^je_j\in\mathfrak{g},\\
 D_\alpha S_2+[S_2,U_\alpha]=(D_\alpha S_2+[S_2,U_\alpha])^je_j.
\ea\end{equation}
Hence, for any smooth $\mathfrak{g}$-valued gauge $S_2([u],\lambda)$ there exists a generalized symmetry $\omega_R^S$ of the ZCC (\ref{2.4}) and the converse statement holds as well. 

Comparing the FG formula for immersion (\ref{4.8}) with the CD immersion formula (\ref{3.5}) we find the gauge
\be
S_2=(\mbox{pr}\omega_R\Phi)\Phi^{-1}.\label{4.9a}
\ee
Hence the FG formula for immersion (\ref{ffg}) is equivalent to the CD immersion formula (\ref{fcd}) for the gauge $S_2$ satisfying (\ref{4.1}).\hfill$\square$

\subsection{The Sym-Tafel immersion formula versus the Fokas-Gel'fand immersion formula}
Under the assumptions of Propositions 1 and 2, we have the following result.

\begin{proposition}
Let $S_1$ and $S_2$ be the two $\mathfrak{g}$-valued matrix functions determined in Propositions 1 and 2, repsectively in terms of a $\lambda$-conformal symmetry and a generalized symmetry of the ZCC (\ref{2.4}) of the LSP associated with an integrable system $\Omega[u]=0$.

If the gauge $S_2$ is a non-singular matrix then there exists a matrix $M=S_1S_2^{-1}$ such that
\be
\beta(\lambda)(D_\lambda\Phi)=M(\mbox{pr}\omega_R\Phi)\label{5.1}.
\ee
The matrix $M$ defines a mapping from the FG immersion formula (\ref{4.8}) to the ST immersion formula (\ref{3.3}).

Alternatively, if the gauge $S_1$ is a non-singular matrix then there exists a matrix  $M^{-1}$ such that
\be
(\mbox{pr}\omega_R\Phi)=M^{-1}\beta(\lambda)(D_\lambda\Phi)\label{5.2}.
\ee
The matrix $M^{-1}$ defines a mapping from the ST immersion formula (\ref{3.3}) to the FG immersion formula (\ref{4.8}).

\end{proposition}

\textbf{Proof.} Equation (\ref{5.1}) or (\ref{5.2}) is obtained by eliminating the wavefunction $\Phi$ from the right-hand side of equations (\ref{3.8}) and \be
\mbox{pr}\omega_R\Phi=S_2\Phi,\label{4.9b}
\ee
 respectively. So the link between the immersion functions $F^{ST}$ and $F^{FG}$ exists, up to a $\g$-valued gauge function.\hfill$\square$

\begin{figure}[h!]
\setlength{\unitlength}{0.14in}
\centering
\begin{picture}(30,12)
\put(1,6){$\Phi\in G$}
\put(1.75,5.5){\vector(3,-1){10.5}}
\put(1.75,7.3){\vector(3,1){9.5}}
\put(4,9.5){$S_1\in\g$}
\put(4,3){$S_2\in\g$}
\put(9.5,11){$F^{ST}=\beta(\lambda)\Phi^{-1}(D_\lambda\Phi)\in\g$}
\put(10.25,1){$F^{FG}=\Phi^{-1}(\mbox{pr}\omega_R\Phi)\in\g$}
\put(10,6){$S_1\circ S_2^{-1}$}
\put(15,2){\vector(0,1){8.5}}
\put(16,10.5){\vector(0,-1){8.5}}
\put(17,6){$S_2\circ S_1^{-1}$}
\end{picture}
\caption{Representation of the relations between the wavefunction $\Phi\in G$\\
 and the $\g$-valued ST and FG formulas for immersions of 2D-soliton surfaces.}
\label{Fig1}
\end{figure}
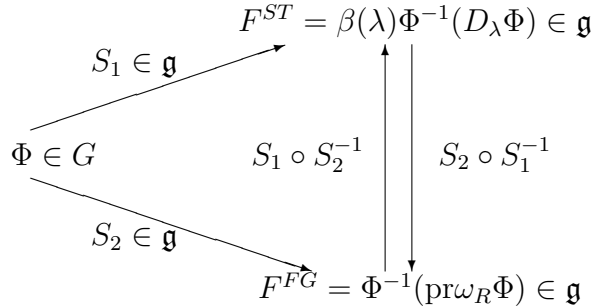

It should be noted that in order to recover soliton surfaces, we have to perform an integration with respect to the curvilinear coordinates in the case of the FG formula. Alternatively, by using the ST immersion formula, we obtain the same soliton surface by differentiating the wavefunction $\Phi$ with respect to the spectral parameter $\lambda$. The connection between the FG and ST approaches for determining the immersion functions $F^{ST}$ and $F^{FG}$ of 2D-surfaces is obtained through the gauge matrix functions $M$ or $M^{-1}$ from the equation (\ref{5.1}) or (\ref{5.2}) respectively (see Fig. \ref{Fig1}).

We can also write direct equations relating the generalized symmetries $\omega_R$ with the Sym-Tafel $\lambda$-conformal symmetry
for the ZCC (\ref{2.4}),  eliminating the gauge $S_2([u],\lambda)$ in (\ref{4.1}) by using (\ref{3.8}). So we get
\be\ba{l}
\beta(\lambda)(D_\lambda\Phi)U_\alpha-\beta(\lambda)\Phi U_\alpha\Phi^{-1}(D_\lambda\Phi)+\beta(\lambda)(D_\lambda U_\alpha)\Phi\\
\hspace{1cm}+\Phi\left[-\mbox{pr}\omega_R(D_\alpha\Phi)+U_\alpha(\mbox{pr}\omega_R\Phi)\right]\Phi^{-1}=0.\label{4.10}
\ea\ee
However, equations (\ref{4.10}) are nonlinear differential equations for the wavefunction $\Phi$, which in general are not easy  to solve.

To conclude, in all three cases we give explicit expressions for 2D-soliton surfaces immersed in the Lie algebra $\g$ and demonstrate that one such surface can be transformed to another one through a gauge.

\section{The sigma model and soliton surfaces}\setcounter{equation}{0}
For the sake of generality we start by considering the general $\mathbb{C}P^{N-1}$ model. 
The problem of constructing integrable surfaces associated with the $\mathbb{C}P^{N-1}$ models and their deformations under various types of dynamics have generated a great deal of interest over the past decades \cite{BBT,MS04,Zakrzewski89}. The most fruitful approach to the study of general properties of this model has been formulated through descriptions of the model in terms of rank-one Hermitian projectors. A matrix $P(z,\bar{z})$ is said to be a rank-one Hermitian projector if
\be
P^2=P,\qquad  P=P^\dagger,\qquad \mbox{tr}P=1.
\ee
The target space of the projector $P$ is determined by a complex line in $\mathbb{C}^N$, i.e. by a one-dimensional vector function $f(z,\bar{z})$ given by 
\be
P=\frac{f\otimes f^\dagger}{f^\dagger f},\label{Pf}
\ee
where $f$ is the mapping $\mathbb{C}\supseteq\Omega\ni z=x+iy\mapsto f=(f_0,f_1,...,f_{N-1})\mathbb{C}^N\backslash\lbrace0\rbrace$. Equation (\ref{Pf}) gives an isomorphism between the equivalence classes of the $\mathbb{C}P^{N-1}$ model and the set of rank-one Hermitian projectors $P$. The equations of motion
\be
\hspace{-2cm}\Omega(P)=[\partial_+\partial_-P,P]=0,\qquad \partial_\pm=\frac{1}{2}(\partial_1\pm i\partial_2),\qquad \partial_1=\partial_{x},~~~\partial_2=\partial_{y}\label{4P}
\ee
and other properties of the model take a compact form when the model is written in terms of the projector.

Now we present some examples which illustrate the theoretical considerations presented in the previous section. Our first example shows that the integrated form of the surface associated with the $\mathbb{C}P^{N-1}$ model admits conformal symmetries which depend on two arbitrary functions of one complex variable.
This model is defined on the Riemann sphere $S^2=\mathbb{C}\cup\lbrace\infty\rbrace$ and its action functional is finite \cite{Zakrzewski89}.  An entire class of solutions of (\ref{4P}) is obtained by acting on the holomorphic (or anti-holomorphic) solution $P$ \cite{GG10} with raising and lowering operators.
These operators are given by
\begin{equation}\ba{l}
\Pi_\pm(P)=\left\lbrace\ba{ll}
\frac{(\partial_\pm P)P(\partial_\mp P)}{\mbox{tr}(\partial_\pm PP\partial_\mp P)} & \mbox{for }(\partial_\pm P)P(\partial_\mp P)\neq0,\\
0 & \mbox{for }(\partial_\pm P)P(\partial_\mp P)=0,
\ea\right.\\
\Pi_-(P_k)=P_{k-1},\qquad \Pi_+(P_k)=P_{k+1}.
\ea\label{recrel}
\end{equation}
The set of $N$ rank-1 projectors $\lbrace P_0,...,P_{N-1}\rbrace$ acts on orthogonal complements of one-dimensional subspaces in $\mathbb{C}^N$ and satisfy the orthogonality and completeness relations
\begin{equation}
P_jP_k=\delta_{jk}P_j,\quad\mbox{(no summation) and} \quad \sum_{j=0}^{N-1} P_j=\mathbb{I}_N,
\end{equation}
where $\mathbb{I}_N$ is the $N\times N$ identity matrix on $\mathbb{C}^N$. These projectors provide a basis of commuting elements in the space of the Hermitian matrices on $\mathbb{C}^N$ and satisfy the Euler-Lagrange equation (written in the form of a conservation law)
\begin{equation}
\partial[\bar{\partial}P_k,P_k]+\bar{\partial}[\partial P_k,P_k]=0,\qquad k=0,1,...,N-1\label{EL2}
\end{equation}
where $\partial=\frac{1}{2}(\partial_x-i\partial_y)$ and $\bar{\partial}=\frac{1}{2}(\partial_x+i\partial_y)$.  For a given set of rank-1 projector solutions $P_k$ of (\ref{EL2}) the $\mathfrak{su}(N)$-valued generalized Weierstrass formula for immersion (GWFI) \cite{Konopelchenko}
\begin{equation}
\hspace{-1cm}F_k(z,\bar{z})=i\int_\gamma(-[\partial P_k,P_k]dz+[\bar{\partial} P_k,P_k]d\bar{z}),\qquad k=0,1,...,N-1
\end{equation}
(where $\gamma$ is a curve locally independent of the trajectory in $\mathbb{C}$) can be explicitly integrated \cite{GG10}
\begin{equation}
F_k(z,\bar{z})=-i\left(P_k+2\sum_{j=0}^{k-1}P_j\right)+\frac{1+2k}{N}\mathbb{I}_N.\label{**}
\end{equation}
The immersion functions $F_k$ satisfy the algebraic conditions
\begin{equation}
\hspace{-2cm}\begin{array}{ll}
~[F_k-ic_k\mathbb{I}_N][F_k-i(c_k-1)\mathbb{I}_N] [F_k-i(c_k-2)\mathbb{I}_N]=0,\qquad& 0<k<N-1\\
~[F_0-ic_0\mathbb{I}_N][F_0-i(c_0-1)\mathbb{I}_N]=0,& k=0,\\
~[F_{N-1}+ic_0\mathbb{I}_N][F_{N-1}+i(c_0-1)\mathbb{I}_N]=0,&k=N-1\\
~\sum_{j=0}^{N-1}(-1)^jF_j=0,\qquad c_k=\frac{1}{N}(1+2k).
\end{array}
\end{equation}
The LSP associated with (\ref{EL2}) is given by \cite{ZM79,Mikhailov86}
\begin{equation}
\partial_\alpha\Phi_k=U_{\alpha k}\Phi_k,\quad U_{\alpha k}=\frac{2}{1\pm\lambda}[\partial_\alpha P_k,P_k],\qquad(U_{1k})^\dagger=-U_{2k},\label{LSP**}
\end{equation}
(where $\alpha=1,2$ stands for $\pm$) with soliton solution $\Phi_k=\Phi_k([P],\lambda)\in SU(N)$ which goes to $\mathbb{I}_N$ as $\lambda\rightarrow\infty$ \cite{ZM79,Zakrzewski89}
\begin{equation}
\begin{array}{l}
\Phi_k=\mathbb{I}_N+\frac{4\lambda}{(1-\lambda)^2}\sum_{j=0}^{k-1}P_j-\frac{2}{1-\lambda}P_k,\qquad \lambda=it,\\
\Phi_k^{-1}=\mathbb{I}_N-\frac{4\lambda}{(1+\lambda)^2}\sum_{j=0}^{k-1}P_j-\frac{2}{1+\lambda}P_k,\qquad t\in\mathbb{R}.
\end{array}
\end{equation}
The recurrence relation (\ref{recrel}) is expressed in terms of rank-1 projectors $P_k$, without any reference to the sequence of functions $f_k$ as in (\ref{Pf}). For the sake of simplicity, in this section, we drop the index $k$ attributed to the $N$ projectors $P_k$.

It is convenient for computational purposes 
to express the $\mathbb{C}P^{N-1}$ model in terms of the matrix
\be
\ba{l}
\theta\equiv i\left(P-\frac{1}{N}\mathbb{I}_N\right)=\theta^se_s\in\mathfrak{su}(N),\\
\left[e_j,e_l\right]=C^s_{jl}e_s,\qquad j,l,s=1,...,N^2-1,
\ea\label{4theta4}
\ee
where $C_{jl}^s$ are the structural constants of $\mathfrak{g}$ and $e_s$ is the basis element for the $\mathfrak{su}(N)$ algebra. Due to the indempotency of the projector $P$ we get the following algebraic restriction on $\theta$:
\be
\theta\cdot\theta=-i\frac{(2-N)}{N}\theta+\frac{(1-N)}{N^2}\mathbb{I}_N\Leftrightarrow P^2=P.\label{4cons}
\ee
The equations of motion in terms of the matrix $\theta$ are
\be
\Omega^j[\theta]=\left[(\partial_1^2+\partial_2^2)\theta,\theta\right]^j=0,\qquad j=1,..., N^2-1\label{EL}
\ee
where $[\cdot,\cdot]^j$ denotes the coefficients of the commutator with respect to the $j^{th}$ basis element $e_j$ for the $\mathfrak{su}(N)$ algebra. The potential matrices $U_\alpha$ in terms of $\theta$ are
\be
\ba{l}
U_1=\frac{-2}{1-\lambda^2}\left([\partial_1\theta,\theta]-i\lambda[\partial_2\theta,\theta]\right),\qquad \lambda=it,\\
U_2=\frac{-2}{1-\lambda^2}\left(i\lambda[\partial_1\theta,\theta]+[\partial_2\theta,\theta]\right),\qquad t\in\mathbb{R}.\label{BB}
\ea
\ee
The wavefunction $\Phi$ in terms of $\theta$ is
\begin{equation}
\ba{l}
\Phi([\theta],\lambda)=\mathbb{I}_N+\frac{4\lambda}{(1-\lambda)^2}\sum_{j=0}^{k-1}\Pi_-^j(\theta)\\
-\frac{2}{1-\lambda}\left(\frac{1}{N}\mathbb{I}_N-i\theta\right)\in SU(N),
\ea
\end{equation}
where $\Pi_\pm$ are the raising and lowering operators acting on the elements $\theta$ of the algebra $\mathfrak{su}(N)$
\begin{equation}
\Pi_-(\theta_k)=\theta_{k-1},\qquad \Pi_+(\theta_k)=\theta_{k+1}.\label{op}
\end{equation}
In what follows, we use the simplified notation of $\Pi_\pm(\theta_k)$ by $\Pi_\pm(\theta)$, where the index $k$ is suppressed.
The operators (\ref{op}) 
are written explicitly as
\begin{equation}
\begin{array}{l}
\Pi_-(\theta)=\frac{\bar{\partial}\theta(\mathcal{E}-i\theta)\partial\theta}{\tr(\bar{\partial}\theta(\mathcal{E}-i\theta)\partial\theta)},\\
\Pi_+(\theta)=\frac{\partial\theta(\mathcal{E}-i\theta)\bar{\partial}\theta}{\tr(\partial\theta(\mathcal{E}-i\theta)\bar{\partial}\theta)},
\end{array}\qquad \mathcal{E}=\frac{1}{N}\mathbb{I}_N,
\end{equation}
where the traces in the denominators are different from zero unless the whole matrix is zero. 
For any functions $f$ and $g$ of one variable, the equations of motion (\ref{EL}) and their LSP (\ref{2.6}) (with the potential matrices (\ref{BB})) admit the conformal symmetries
\be
\omega_{C_i}=\left[f(x)\partial_1\theta^j+g(y)\partial_2\theta^j\right]\frac{\partial}{\partial\theta^j},\qquad i=1,2.
\ee
The vector fields $\omega_{C_i}$ are related to the fields $\eta_{c_i}$ defined on the jet space $\mathcal{M}=[(\Phi,U_\alpha)]$
\be
\eta_{C_i}=(\partial_i\Phi^j)\frac{\partial}{\partial\Phi^j}+(\partial_iU_\alpha^j)\frac{\partial}{\partial U_\alpha^j},\qquad i=1,2
\ee
which are conformal symmetries of the LSP (\ref{2.6}).
The integrated form of the surface is given by the FG formula \cite{GPR14}
\be
F^{FG}=\Phi^{-1}\left(f(x)U_1+g(y)U_2\right)\Phi\in\mathfrak{su}(N).
\ee

\subsection{Soliton surfaces associated with the $\mathbb{C}P^1$ sigma model}
We give a simple example to illustrate the construction of 2D-soliton surfaces associated with the $\mathbb{C}P^1$ model ($N=2$) introduced in previous section. The only solutions with finite action of the $\mathbb{C}P^1$ model are holomorphic and antiholomorphic projectors \cite{Zakrzewski89}. 
The rank-one Hermitian projectors (i.e. holomorphic $P_0$ and antiholomorphic $P_1$) based on the Veronese sequence $f_0=(1,z)$, take the form
\begin{equation}
\begin{array}{ll}
P_0=\frac{f_0\otimes f_0^\dagger}{f_0^\dagger f_0}=\frac{1}{1+\vert z\vert^2}\left(\begin{array}{cc}
1 & \bar{z} \\
z & \vert z\vert^2
\end{array}\right),& k=0\\
P_1=\frac{f_1\otimes f_1^\dagger}{f_1^\dagger f_1}=\frac{1}{1+\vert z\vert^2}\left(\begin{array}{cc}
\vert z\vert^2 & -\bar{z} \\
-z & 1
\end{array}\right),& k=1
\end{array}\label{megastar}
\end{equation}
where $f_1=(\mathbb{I}_2-P_0)\partial f_0$. The corresponding integrated forms of the surfaces are given by the GWFI (\ref{**})
\begin{equation}
\begin{array}{l}
F_0=i(\frac{1}{2}\mathbb{I}_2-P_0)=\frac{i}{1+\vert z\vert^2}\left(\begin{array}{cc}
\frac{1}{2}(\vert z\vert^2-1) & -\bar{z} \\
-z & \frac{1}{2}(1-\vert z\vert^2)
\end{array}\right)\in\mathfrak{su}(2),\\
~\\
F_1=-i(P_1+2P_0)+\frac{3i}{2}\mathbb{I}_2=F_0.
\end{array}
\end{equation}
From equation (\ref{LSP**}) the potential matrices $U_{\alpha k}$ become
\begin{equation}
\begin{array}{l}
U_{10}=U_{11}=\frac{2}{(\lambda+1)(1+\vert z\vert^2)^2}\left(\begin{array}{cc}
-\bar{z} & -\bar{z}^2\\
1 & \bar{z}
\end{array}\right),\\
U_{20}=U_{21}=\frac{2}{(\lambda-1)(1+\vert z\vert^2)^2}\left(\begin{array}{cc}
-z & 1 \\
-z^2 & z
\end{array}\right),\\
\lambda=it,\qquad t\in\mathbb{R}.
\end{array}
\end{equation}
The $SU(2)$-valued soliton wavefunctions $\Phi_k$ in the LSP (\ref{LSP**}) for the $\mathbb{C}P^1$ model have the form
\begin{equation}
\begin{array}{l}
\Phi_0=\frac{1}{1+\vert z\vert^2}\left(\begin{array}{cc}
\frac{-i+t+(i+t)\vert z\vert^2}{t-i} & \frac{-2i\bar{z}}{t-i} \\
\frac{-2iz}{t+i} & \frac{i+t+(t-i)\vert z\vert^2}{t+i}
\end{array}\right),\\
\Phi_1=\frac{1}{1+\vert z\vert^2}\left(\begin{array}{cc}
\frac{1+t^2+(t+i)^2\vert z\vert^2}{(t-i)^2} & \frac{2(1-it)\bar{z}}{(t-i)^2} \\
\frac{-2i(t-i)z}{(t+i)^2} & \frac{1+t^2+(t-i)^2\vert z\vert^2}{(t+i)^2}
\end{array}\right).
\end{array}
\end{equation}

Let us now consider separately four different analytic descriptions for the immersion functions of 2D-soliton surfaces in Lie algebras which are related to four different types of symmetries.

\textbf{1.} The ZCC (\ref{2.4}) of the $\mathbb{C}P^1$ model admits a conformal symmetry in the spectral parameter $\lambda$. The tangent vectors $D_\alpha F_k^{ST}$ associated with this symmetry are given by
\begin{equation*}
D_\alpha F_k^{ST}=\Phi_k^{-1}(D_\lambda U_{\alpha k})\Phi_k,\qquad \alpha,k=1,2
\end{equation*}
and are linearly independent. The integrated forms of the 2D-surfaces in $\mathfrak{su}(2)$ are given by the ST formulas (\ref{fst})
\begin{equation*}
\hspace{-2.5cm}\begin{array}{l}
\vspace{3mm}F_0^{ST}=\Phi_0^{-1}(D_\lambda\Phi_0)=\frac{2i}{(1+t^2)^2(1+\vert z\vert^2)^2}\\
\vspace{3mm}\left(\begin{array}{cc}
-\vert z\vert^2[t^2-3+\vert z\vert^2(1+t^2)] & \bar{z}[(t+i)^2+\vert z\vert^2(3+2it+t^2)] \\
z[(t-i)^2+\vert z\vert^2(3-2it+t^2)] & \vert z\vert^2[t^2-3+\vert z\vert^2(t^2+1)]
\end{array}\right),\qquad k=0\\
~\\
\vspace{3mm}F_1^{ST}=\Phi_1^{-1}(D_\lambda\Phi_1)=\frac{2i}{(1+t^2)^2(1+\vert z\vert^2)^2}\\
\vspace{3mm}\left(\begin{array}{cc}
-[(t^2+1)(1+2\vert z\vert^4)+3\vert z\vert^2(t^2-3)] & \bar{z}[6it-5+t^2+\vert z\vert^2(7+6it+t^2)] \\
z[t^2-6it-5+\vert z\vert^2(7+t^2-6it)]  & (t^2+1)(1+2\vert z\vert^4)+3\vert z\vert^2(t^2-3)
\end{array}\right),\quad k=1
\end{array}\label{F0st}
\end{equation*}
where, without loss of generality, we can put $\beta(\lambda)=1$ in the expression (\ref{fst}).
The surfaces $F_k^{ST}$ satisfy 
\begin{equation*}
(F^{ST}_k)^2+\frac{1}{4}\mathbb{I}_2=0,
\end{equation*}
for $k=0,1$ and have positive constant Gaussian and mean curvatures. Hence, they are spheres (see Fig. 2a)
\begin{equation}
K_0=K_1=4,\qquad H_0=H_1=4.\label{1KH}
\end{equation}
The surfaces $F^{ST}_k$ in Cartesian coordinates $(x,y)$ take the form for $z=x+iy$
\be
F_k^{ST}=\left\lbrace\frac{x}{1+x^2+y^2},\frac{y}{1+x^2+y^2},\frac{1-x^2-y^2}{2(1+x^2+y^2)}\right\rbrace
\ee
The $\mathfrak{su}(2)$-valued gauges $S_k^{ST}$ associated with the ST immersion functions $F_k^{ST}$ (\ref{F0st}) take the form
\begin{equation}
\begin{array}{l}
S_0^{ST}=(D_\lambda\Phi_0)\Phi_0^{-1}=\frac{2i}{1+\vert z\vert^2}\left(\begin{array}{cc}
\frac{-\vert z\vert^2}{t^2+1} & \frac{\bar{z}}{(t-i)^2} \\
\frac{z}{(t+i)^2} & \frac{\vert z\vert^2}{t^2+1}
\end{array}\right), \\
S_1^{ST}=(D_\lambda\Phi_1)\Phi_1^{-1}=\frac{2i}{1+\vert z\vert^2}\left(\begin{array}{cc}
\frac{-(1+2\vert z\vert^2)}{t^2+1} & \frac{\bar{z}(t+i)^2}{(t-i)^4} \\
\frac{z(t-i)^2}{(t+i)^4} & \frac{1+2\vert z\vert^2}{t^2+1}
\end{array}\right),\\
\det S_k^{ST}\neq0,\qquad \tr S_k^{ST}=0.
\end{array}
\end{equation}

\textbf{2.} The surfaces $F^{g}_k\in\mathfrak{su}(2)$ associated with the scaling symmetries of the ZCC (\ref{2.4}) associated with equations of the $\mathbb{C}P^1$ model (\ref{EL2})
\begin{equation}
\omega^{g}_k=\left(D_1(zU_{1k})+\bar{z}(D_2U_{1k})\right)\frac{\partial}{\partial\theta^1}+\left(z(D_1U_{2k})+D_2(\bar{z}U_{2k})\right)\frac{\partial}{\partial\theta^2},
\end{equation}
have the integrated form \cite{GP12conf}
\begin{equation}
F_k^g=\Phi^{-1}_k(zU_{1k}+\bar{z}U_{2k})\Phi_k,\qquad k=0,1
\end{equation}
where $\theta^1$ and $\theta^2$ are complex-valued functions determined from the $\mathfrak{su}(2)$ Lie algebra (\ref{4theta4}). The surfaces $F_k^g$ also have constant positive curvatures
\begin{equation}
K_0=K_1=-4\lambda^2,\qquad H_0=H_1=-4i\lambda,\qquad i\lambda\in\mathbb{R}.
\end{equation}
The surfaces $F_k^g$ are not spheres (as in the previous cases (\ref{1KH})) since they have boundaries (see Fig. 2b). The surfaces $F^g_k$ can be given in the parametric form
\be
\hspace{-1cm}\ba{l}
F_k^g=\left(\frac{x^3-2x^2y+x(y^2-1)-2y(1+y^2)}{(1+x^2+y^2)^2},-\frac{2x^3+x^2y+y(y^2-1)+2x(1+y^2)}{(1+x^2+y^2)^2},\frac{2(x^2+y^2)}{(1+x^2+y^2)^2}\right)
\ea
\ee
The  $\mathfrak{su}(2)$-valued gauges $S_k^g=($pr$\omega^g\Phi_k)\Phi_k^{-1}$ associated with the scaling symmetries $\omega^g_k$ are given by
\begin{equation}
\hspace{-2.5cm}S_0^g=S_1^g=\frac{2}{(t^2+1)(1+\vert z\vert^2)^2}\left(\begin{array}{cc}
2it\vert z\vert^2 & i\bar{z}[i-t+\vert z\vert^2(t+i)] \\
z[1-it+\vert z\vert^2(1+it)] & -2it\vert z\vert^2
\end{array}\right),
\end{equation}
where det$S_k^g\neq0$.

\textbf{3.} In the case of surfaces associated with the conformal symmetries 
\begin{equation}
\omega^c_k=-g_k(z)\partial-\bar{g}_k(\bar{z})\bar{\partial}\label{etoile},
\end{equation}
where for simplicity we have assumed that $g_k(z)=1+i$, the $\mathfrak{su}(2)$-valued immersion functions $F^c_k$ are given by \cite{GP11}
\begin{equation}
F^c_k=\Phi_k^{-1}(U_{1k}+U_{2k})\Phi_k,
\end{equation}
where
\begin{equation}
\begin{array}{l}
U_{10}+U_{20}=U_{11}+U_{21}=\frac{2}{(t^2+1)(1+\vert z\vert^2)^2}\\
\left(\begin{array}{cc}
2z+i(t+i)(z+\bar{z}) & -1-it+i\bar{z}^2(t+i) \\
1+z^2+it(z^2-1) & -[2z+i(t+i)(z+\bar{z})]
\end{array}\right).\\
\end{array}
\end{equation}
By further computation it can be verified that the Gaussian curvature and the mean curvature corresponding to the surfaces $F^c_k$ are not constant for any value of $\lambda$. The fact that the surfaces $F^c_k$ have the Euler-Poincar\'{e} characters \cite{GG10}
\be
\chi_k=\frac{-1}{\pi}\int\int_{S^2}\partial\bar{\partial}\ln{[tr(\partial P_k\cdot\bar{\partial}P_k)]}dx^1dx^2
\ee
equal to 2 and positive Gaussian curvature $K>0$ means that the surfaces $F^c_k$ are homeomorphic to ovaloids (see Fig. 2c). The surfaces $F_k^c$ associated with the conformal symmetries $\omega_k^c$ are cardioid surfaces which can be parametrized as follows
\be
\hspace{-1cm}F_k^c=\left(\frac{x^2-1-4xy-y^2}{(1+x^2+y^2)^2},-\frac{2(1+x^2+xy-y^2)}{(1+x^2+y^2)^2},\frac{2(2x-y)}{(1+x^2+y^2)^2}\right)
\ee

The $\mathfrak{su}(2)$-valued gauges $S_k^c$ associated with the conformal symmetries $\omega^c_k$
take the form
\begin{equation}
\hspace{-1cm}\begin{array}{l}
\vspace{3mm}S_0^c=(\mbox{pr}\omega^c\Phi_0)\Phi^{-1}_0=\frac{2}{(1+\vert z\vert^2)^2}\\
\vspace{3mm}\left(\begin{array}{cc}
\frac{-i(1-i)(t-i)z+(1+i)(1-it)\bar{z}}{t^2+1} & \frac{(1-i)(1+it)+(1+i)(1-it)\bar{z}^2}{(t-i)^2} \\
\frac{i(1+i)(t+i)-(1-i)z^2(t-i)}{(t+i)^2} & \frac{(1-i)(1+it)z+i(1+i)(t+i)\bar{z}}{t^2+1}
\end{array}\right),\qquad k=0\\
~\\
\vspace{3mm}S_1^c=(\mbox{pr}\omega^c\Phi_1)\Phi^{-1}_1=\frac{2}{(1+\vert z\vert^2)^2}\\
\vspace{3mm}\left(\begin{array}{cc}
\frac{-i(1-i)(t-i)z+(1+i)(1-it)\bar{z}}{t^2+1} & \frac{(t+i)^2[(1-i)(1+it)+(1+i)(1-it)\bar{z}^2]}{(t-i)^4} \\
\frac{i(t-i)^2[(1+i)(t+i)-(1-i)(t-i)z^2]}{(t+i)^4} & \frac{(1-i)(1+it)z+i(1+i)(t+i)\bar{z}}{t^2+1}
\end{array}\right),\qquad k=1
\end{array}
\end{equation}
where det$S_k^c\neq0$.

\textbf{4.} If the generalized symmetries of the $\mathbb{C}P^1$ model (\ref{EL2}) are written in the evolutionary form
\begin{equation}
\hspace{-1cm}\begin{array}{l}
\omega^R_k=(D_1^2U_{1k}+D_2^2U_{1k}+[D_1U_{1k},U_{1k}]+[D_2U_{1k},U_{1k}])\frac{\partial}{\partial\theta^1}\\
\hspace{1cm}+(D_1^2U_{2k}+D_2^2U_{2k}+[D_2U_{2k},U_{2k}]+[D_1U_{2k},U_{2k}])\frac{\partial}{\partial\theta^2},\quad k=1,2
\end{array}
\label{omegaQ}
\end{equation}
(where $\theta^1$ and $\theta^2$ are complex-valued functions obtained from (\ref{4theta4})) 
then the $\mathfrak{su}(2)$-valued integrated form of the immersion becomes \cite{GP12conf}
\begin{equation}
F^{FG}_k=\Phi^{-1}(\mbox{pr}\omega^R_k\Phi_k)=\Phi_k^{-1}(D_1U_{1k}+D_2U_{2k})\Phi_k. \label{FGequation1}
\end{equation}
The tangent vectors to this surface are given by
\begin{equation*}
\hspace{-2.7cm}\begin{array}{l}
D_1F_k^{FG}=\Phi_k^{-1}(\mbox{pr}\omega^R_kU_{1k})\Phi_k=\Phi_k^{-1}(D_1^2U_{1k}+D_2^2U_{1k}+[D_1U_{1k},U_{1k}]+[D_2U_{1k},U_{1k}])\Phi_k,\\
\\
D_2F_k^{FG}=\Phi_k^{-1}(\mbox{pr}\omega^R_kU_{2k})\Phi_k=\Phi_k^{-1}(D_1^2U_{2k}+D_2^2U_{2k}+[D_2U_{2k},U_{2k}]+[D_1U_{2k},U_{2k}])\Phi_k.
\end{array}
\end{equation*}
The surfaces $F_k^{FG}$ also have positive Gaussian curvatures $K>0$ and the Euler-Poincar\'e characters are equal to 2. In the parametrization $x,y$, the surfaces $F_k^{FG}$ take the form
\be
\hspace{-2.5cm}\ba{c}
F_k^{FG}=\left(-\frac{x^3-6x^2y-x(1+3y^2)+2y(1+y^2)}{(1+x^2+y^2)^3},\frac{2x^3+y+3x^2y-y^3+x(2-6y^2)}{(1+x^2+y^2)^3},-\frac{2(x^2-4xy-y^2)}{(1+x^2+y^2)^3}\right)
\ea
\ee
and they are homeomorphic to ovaloids.

The $\mathfrak{su}(2)$-valued gauges $S_k^{FG}$ associated with the generalized symmetry $\omega^R_k$ take the form
\begin{equation}
S_k^{FG}=(\mbox{pr}\omega^R_k\Phi_k)\Phi_k^{-1}=D_1U_{1k}+D_2U_{2k}.
\end{equation}
Under the assumption that $P_k$ are holomorphic or antiholomorphic projectors (\ref{megastar}) the gauges $S_k^{FG}$ take the form
\begin{equation}
\hspace{-2.7cm}\begin{array}{l}
S_0^{FG}=S_1^{FG}=\frac{4}{(t^2+1)(1+\vert z\vert^2)^3}\left(\begin{array}{cc}
-z^2(1+it)+\bar{z}^2(1-it) & \bar{z}^3(1-it)+z(it+1) \\
-iz^3(t-i)+i\bar{z}(t+i) & z^2(1+it)-\bar{z}^2(1-it)
\end{array}\right),\\
~\\

\end{array}
\end{equation}
where $\det S_k^{FG}\neq0$. Hence the mappings $M_k=S_k^{ST}(S_k^{FG})^{-1}$ from the FG immersion formulas to the ST immersion formulas are given by
\begin{equation}
\hspace{-2.5cm}\begin{array}{l}
M_0=S_0^{ST}(S_0^{FG})^{-1}=\frac{1}{2(t^2+1)}\\
\left(\begin{array}{cc}
\frac{-2iz^3(t-i)+\bar{z}[(t+i)^2+\vert z\vert^2(t^2+1)]}{z(t-i)} & \frac{z[(t+i)^2+(t^2+1)\vert z\vert^2+2(1+it)]}{t-i} \\
-\frac{[z^3(1+t^2)+2\bar{z}(1-it)+z(t-i)^2]}{t+i} & \frac{z(t-i)^2+\vert z\vert^2z(t^2+1)+2i\bar{z}^3(t+i)}{\bar{z}(t+i)}
\end{array}\right),\\
~\\
M_1=S_1^{ST}(S_1^{FG})^{-1}=\frac{i}{2\vert z\vert^2}\\
\left(\begin{array}{cc}
\frac{-(1+2\vert z\vert^2)}{t^2+1} & \frac{(i+t)^2\bar{z}}{(t-i)^4} \\
\frac{z(t-i)^2}{(t+i)^4} & \frac{1+2\vert z\vert^2}{t^2+1}
\end{array}\right)\left(\begin{array}{cc}
z^2(it+1)+\bar{z}^2(it-1) & -z(it+1)+\bar{z}^3(it-1)\\
z^3(it+1)+(1-it)\bar{z} & -z^2(1+it)+(1-it)\bar{z}^2
\end{array}\right),
\end{array}
\end{equation}
where det$M_k\neq0$. 
Conversely, the gauges $M_k^{-1}=S_k^{FG}(S_k^{ST})^{-1}$ do exist. Hence there exist mappings from the ST immersion formulas to the FG immersion formulas
\begin{equation}
M_0^{-1}=S_0^{FG}(S_0^{ST})^{-1}=\frac{2}{(1+\vert z\vert^2)^3}\left(\hspace{-2mm}\begin{array}{cc}
m_{11} & \hspace{-1mm}m_{12} \\
m_{21} & \hspace{-1mm}m_{22}
\end{array}\hspace{-2mm}\right)\hspace{-1mm},
\end{equation}
where
\begin{equation}
\begin{array}{l}
\vspace{2mm}m_{11}=\frac{(t-i)^2z+(1+t^2)\vert z\vert^2 z+2(it-1)\bar{z}^3}{(i+t)\bar{z}},\\
\vspace{2mm}m_{12}=\frac{2(1+it)z^2+(i+t)^2\bar{z}^2+(1+t^2)\vert z\vert^2\bar{z}^2}{(i-t)z},\\
\vspace{2mm}m_{21}=\frac{(t-i)^2z^2+(1+t^2)\vert z\vert^2z^2+2(1-it)\bar{z}^2}{(i+t)\bar{z}},\\
m_{22}=\frac{-2(1+it)z^3+(i+t)^2\bar{z}+(1+t^2)\vert z\vert^2 \bar{z}}{(t-i)z}.
\end{array}\nonumber
\end{equation}
\begin{equation}
\begin{array}{l}
\vspace{2mm}M_1^{-1}=S_1^{FG}(S_1^{ST})^{-1}=\frac{2}{(t^2+1)(1+\vert z\vert^2)^3(1+4\vert z\vert^2)}\\
\vspace{2mm}\left(\begin{array}{cc}
-z^2(1+it)+\bar{z}^2(1-it) & \bar{z}^3(1-it)+z(1+it) \\
-z^3(1+it)+\bar{z}(it-1) & z^2(1+it)-\bar{z}^2(1-it)
\end{array}\right)\cdot\\
\left(\hspace{-1mm}\begin{array}{cc}
(1+2\vert z\vert^2)(1+it)(i+t) & \frac{-i\bar{z}(i+t)^4}{(t-i)^2} \\
\frac{-iz(t-i)^4}{(i+t)^2} & \hspace{-5mm}(1+2\vert z\vert^2)(1-it)(-i+t)
\end{array}\hspace{-1mm}\right)
\end{array}
\end{equation}

\begin{figure}[h!]
\setlength{\unitlength}{0.14in}
\centering
\includegraphics[scale=0.5]{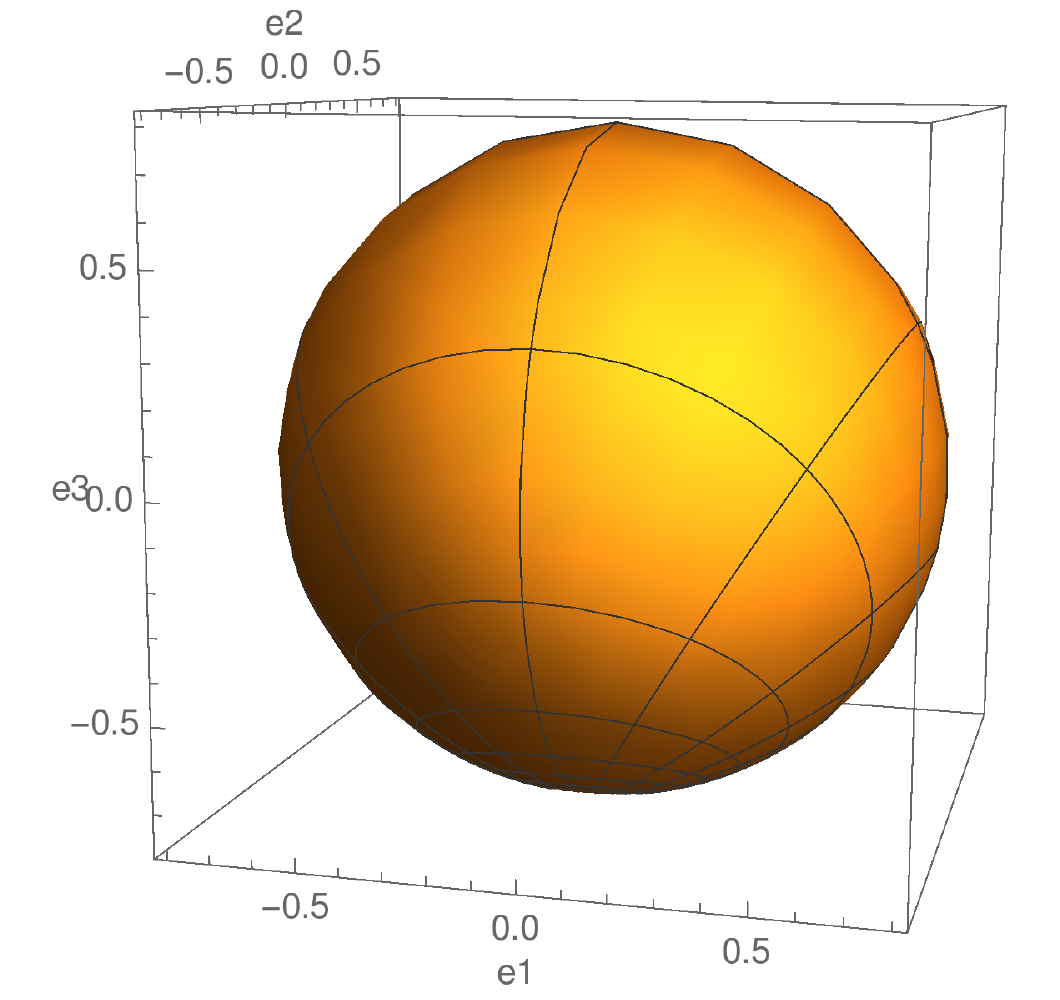}\hspace{5mm}
\includegraphics[scale=0.5]{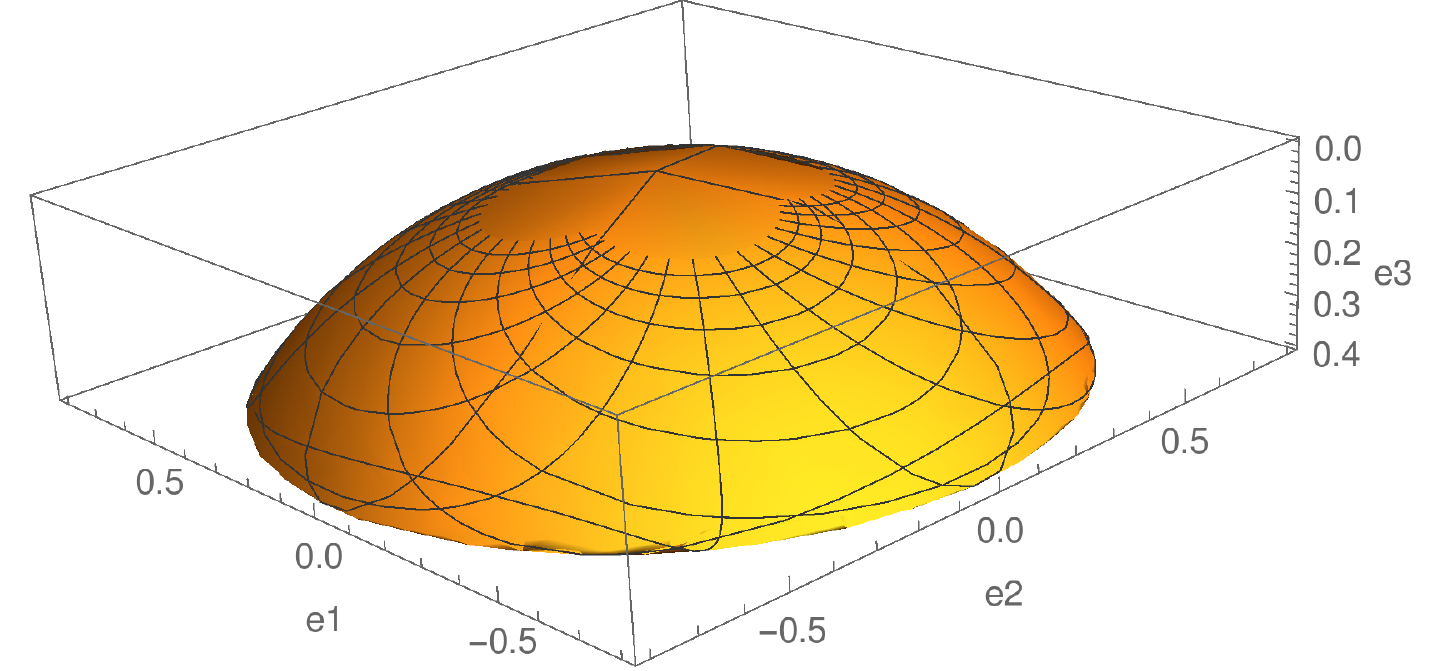}\hspace{5mm}
\begin{picture}(32,1)
\put(5,0){(a)}
\put(23,0){(b)}
\end{picture}
\includegraphics[scale=0.5]{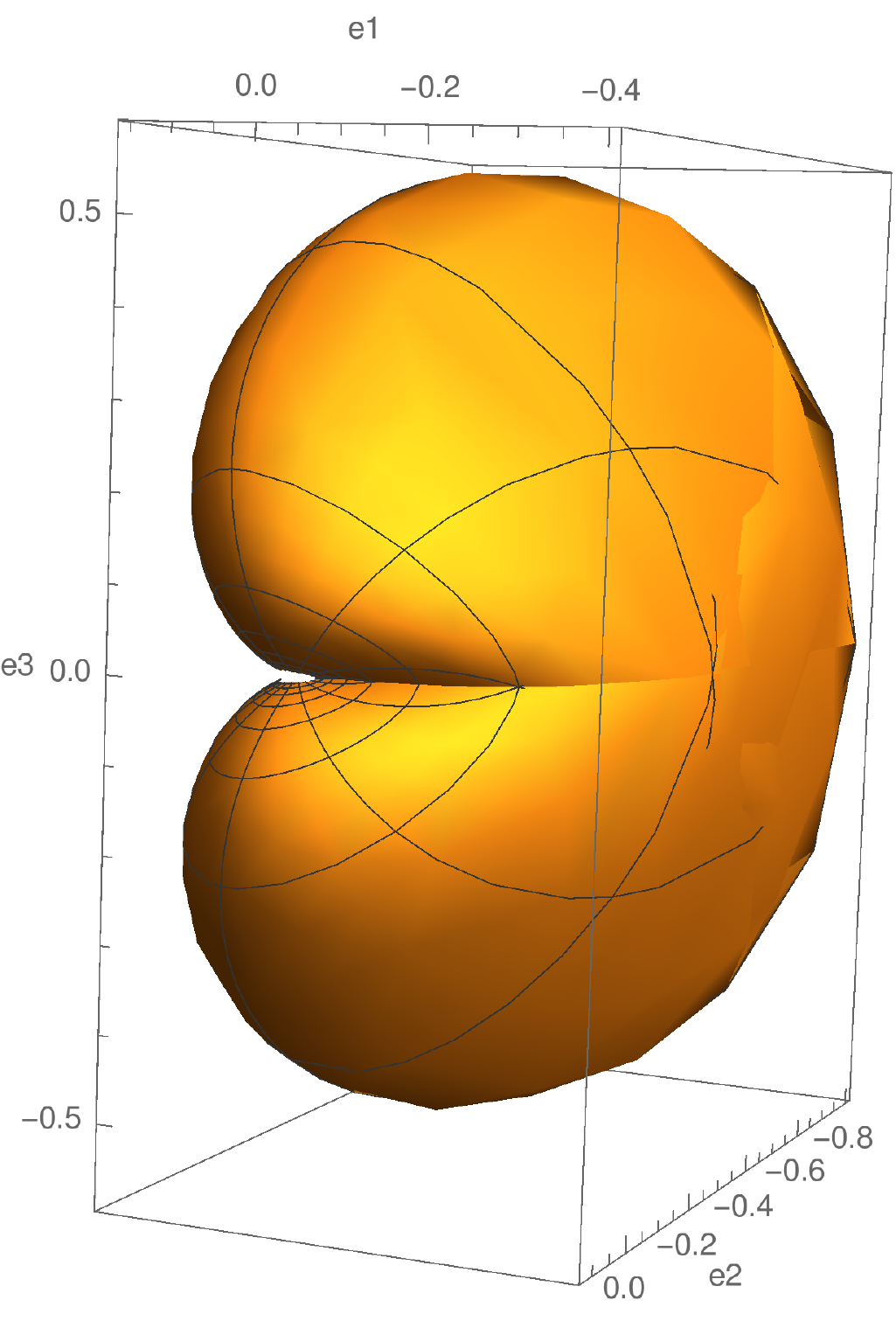}\hspace{5mm}
\includegraphics[scale=0.5]{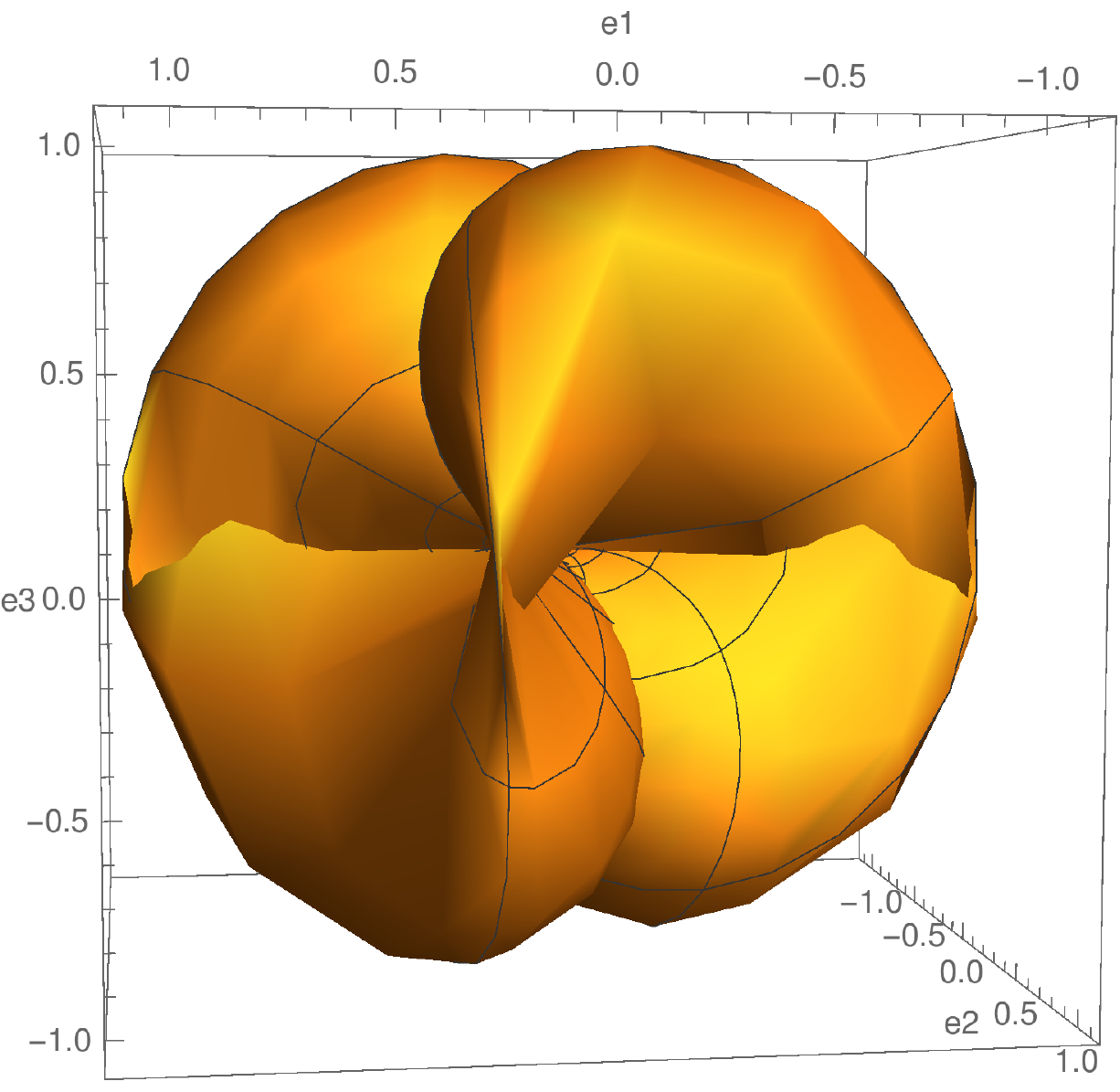}
\begin{picture}(32,1)
\put(5,0){(c)}
\put(23,0){(d)}
\end{picture}
\caption{Surfaces $F^{ST}_0$ in (a), $F^g_0$ in (b), $F^c_0$ in (c) and $F^{FG}_0$ in (d) for $k=0,1$ and $\lambda=i/2$, and $\xi_\pm=x\pm i y$ with $x,y\in[-5,5]$. The axes indicate the components of the immersion function in the basis for $\mathfrak{su}(2)$, \\$e_1=\left(\hspace{-2mm}\ba{cc}0&i\\ i&0 \ea\hspace{-2mm}\right),\; e_2=\left(\hspace{-2mm}\ba{cc}0&-1\\ 1&0 \ea\hspace{-2mm}\right),\; e_3=\left(\hspace{-2mm}\ba{cc}i&0\\ 0&-i \ea\hspace{-2mm}\right).$}
\end{figure}

\section{Concluding remarks}\setcounter{equation}{0}
In this paper we have shown how three different analytic descriptions for the immersion function of 2D-soliton surfaces can be related through different $\mathfrak{g}$-valued gauge transformations. The existence of such gauges is demonstrated by reducing the problem to that of mappings between different forms of the immersion formulas for three types of symmetries : conformal transformations in the spectral parameter, gauge symmetries of the LSP and generalized symmetries of the integrable systems. We have investigated the geometric consequences of these mappings and rephrased them as requirements for the existence of the corresponding vector fields and their prolongations acting on a solution $\Phi$ of the associated LSP for an integrable PDE. The explicit expressions for these relations, which we have established, have provided us with a tool for distinguishing between the cases in which soliton surfaces can be or cannot be related 
 among them, see Proposition 3.

The task of finding an increasing number of soliton surfaces associated with integrable systems is related to the symmetry properties of these systems. 
The construction of soliton surfaces started with the contribution of Sym \cite{Sym82,Sym85} and Tafel \cite{Tafel95} providing a formula for the immersion of integrable surfaces which are extensively used in the literature (see e.g. \cite{Bobenko,Cieslinski97,Cieslinski07,DS92,FG96,FGFL,GG10,Grundland15,GP11,GP12,GP12conf,GPR14,GY09,RS00,Sym82,Sym85,Tafel95}). 
In this paper we have addressed the question and formulated easily verifiable conditions which ensure that the ST formula produces a desired result. This advance can assist future studies of 2D-soliton surfaces with integrable models, which can describe more diverse types of surfaces than the ones discussed in three-dimensional Euclidean space for the $\mathbb{C}P^1$ sigma model. It may be worthwhile to extend the investigation of soliton surfaces to the case of the sigma models defined on other homogeneous spaces via Grassmann models and possibly to models associated with octonion geometry. This case could lead to different classes and types of surfaces than those studied in this paper. This task will be explored in a future work.

\section*{Acknowledgements}
AMG has been partially supported by a research grant from NSERC of Canada and would also like to thank the Dipartimento di Mathematica e Fisica of the Universit\'a Roma Tre and the Dipartimento di Mathematica e Fisica of Universit\'a del Salento for its warm hospitality.\\ 
DL  has been partly supported by the Italian Ministry of Education and Research, 2010 PRIN {\it Continuous and discrete nonlinear integrable evolutions: from water waves to symplectic maps}.\\
LM has been partly supported by the Italian Ministry of Education and Research, 2011 PRIN  {\it Teorie geometriche e analitiche dei sistemi Hamiltoniani in dimensioni finite e infinite}.\\ DL and LM are also supported by INFN   IS-CSN4 {\it Mathematical Methods of Nonlinear Physics}.7890.

\bibliographystyle{actapoly}
\bibliography{biblio}

\section*{References}
\begin{thebibliography}{10}

\bibitem{BBT}
Babelon O, Bernard D and Talon M 2006 Introduction to Classical Integrable Systems (Cambridge Monographs on Mathematical Physics) (Cambridge: Cambridge University Press)  DOI: 10.1017/CBO9780511535024
 
\bibitem{Bobenko}
Bobenko AI (1994) Surfaces in terms of 2 by 2 matrices. Old and new integrable cases in Harmonic Maps and Integrable Systems, Eds Fordy A, Wood J (Braunschwieg, Vieweg) DOI: 10.1007/978-3-663-14092-4\_5

\bibitem{cn91}
Calogero F and Nucci MC, Lax pairs galore, {\em J. Math. Phys.}, {\bf 32}  72--74 (1991) DOI: 10.1063/1.529096

\bibitem{Cartan53}
Cartan E 1953 Sur la Structure des Groupes Infinis de Transformation Chapitre I. Les Syst\`emes Diff\'erentiels en Involution (Paris, Gauthier-Villars).

\bibitem{Cieslinski97}
Cie\'{s}li\'{n}ski J 1997 A generalized formula for integrable classes of surfaces in {L}ie algebras \emph{Journal of Mathematical Physics} \textbf{38} 4255--4272 \\DOI: 10.1063/1.532093

\bibitem{Cieslinski07}
Cieslinski J 2007 Pseudospherical surfaces on time scales: a geometric deformation and the spectral approach J. Phys. A: Math. Theor. 40 12525-38 \\DOI: 10.1088/1751-8113/40/42/S02

\bibitem{DS92}
Doliwa A and Sym A 1992  Constant mean curvature surfaces in $E^3$ as an example of soliton surfaces, \textit{Nonlinear Evolution Equations and Dynamical Systems}  (Boiti M, Martina L and Pempinelli F, eds, World Scientific, Singapore, pp 111-7)

\bibitem{FG96}
Fokas A~S and Gel'fand I~M 1996  Surfaces on {L}ie groups, on {L}ie algebras,
and their integrability {\em Comm. Math. Phys.} \textbf{177} 203--220

\bibitem{FGFL}
Fokas A~S, Gel'fand I~M, Finkel F and Liu Q~M 2000 A formula for constructing  infinitely many surfaces on {L}ie algebras and integrable equations  \emph{Sel. Math.} \textbf{6} 347--375 DOI: 10.1007/PL00001392

\bibitem{GG10}
Goldstein P~P and Grundland A~M 2010 Invariant recurrence relations for $CP^{N-1}$ models J. Phys. A: Math Theor. \textbf{43}, 265206 (18pp) \\DOI: 10.1088/1751-8113/43/26/265206

\bibitem{Grundland15}
Grundland AM 2016 Soliton surfaces in generalized symmetry approach J. Theor. Math. Phys. (accepted.)

\bibitem{GP11}
 Grundland A M  and Post S 2011 Soliton surfaces associated with generalized symmetries of integrable equations {\em J. Phys. A.: Math. Theor.}{\bf 44} 165203 (31pp) DOI: 10.1088/1751-8113/44/16/165203

\bibitem{GP12}
Grundland A~M and Post S 2012 Surfaces immersed in Lie algebras associated with elliptic integrals J Phys A: Math Theor \textbf{45}, 015204 (20pp).\\ DOI: 10.1088/1751-8113/45/1/015204

\bibitem{GP12conf}
Grundland AM and Post S 2012 Soliton surfaces associated with $\mathbb{C}P^{N-1}$ sigma models J. Phys. Conf. Series 380, 012023 pp 1-14\\ DOI: 10.1088/1742-6596/380/1/012023

\bibitem{GPR14}
Grundland AM, Post S and Riglioni D 2014 Soliton surfaces and generalized symmetries of integrable systems {\em J. Phys. A.: Math. Theor.}{\bf 47} 015201 (14pp) DOI: 10.1088/1751-8113/47/1/015201

\bibitem{GY09}
Grundland AM and Yurdusen I 2009 On analytic descriptions of two-dimensional surfaces associated with the $\mathbb{C}P^{N-1}$ sigma model, J. Phys. A: Math. Theor. 42 172001 DOI: 10.1088/1751-8113/42/17/172001

\bibitem{glsgal}
Gubbiotti G, Scimiterna C and Levi D 2016 Linearizability and fake Lax pair for a consistent around the cube nonlinear non--autonomous quad--graph equation, {\it Teor. Math. Phys.}, in press.

\bibitem{hb} 
Hay M and Butler S, Simple identification of fake Lax pair, arXiv:1311.2406v1.

\bibitem{Helein01}
Helein F 2001 Constant Mean Curvature Surfaces, Harmonic Maps and Integrable Systems (Lectures in Mathematics) (Boston, MA: Birkhauser). \\DOI: 10.1007/978-3-0348-8330-6

\bibitem{Konopelchenko}
Konopelchenko B G, 1996 Induced surfaces and their integrable dynamics. Stud. Appl. Math. \textbf{96}, 9--51. \\DOI: 10.1002/sapm19969619

\bibitem{lsz90}
Levi D, Sym A and Tu GZ, A working algorithm to isolate integrable surfaces in $E^3$, preprint DF INFN 761, Roma Oct. 10$^{th}$, 1990. \\DOI: 10.1016/0375-9601(90)90897-W

\bibitem{lll10}
Li YQ, Li B and Lou SY, Constraints for Evolution Equations with Some Special Forms of Lax Pairs and Distinguishing Lax Pairs by Available Constraints, arXiv:1008.1375v2 [nlin.SI].

\bibitem{MS04}
Manton N and Sutcliffe P 2004 Topological Solitons (Cambridge Monographs on Mathematical Physics) (Cambridge: Cambridge University Press) \\DOI: 10.1017/CBO9780511617034

\bibitem{m02}
Marvan M 2002 On the Horizontal Gauge Cohomology and Nonremovability of the Spectral Parameter, {\em Acta Appl. Math.} {\bf 72}  51--65. \\DOI: 10.1023/A:1015218422059

\bibitem{m04}
Marvan M 2004 Reducibility of zero curvature representations with application to recursion operators, {\em Acta Appl. Math.}  {\bf 83}  39--68. \\DOI: 10.1023/B:ACAP.0000035588.67805.0b

\bibitem{m10}
Marvan M 2010 On the spectral parameter problem, {\em Acta Appl. Math.} {\bf 109}  239--255. \\DOI: 10.1007/s10440-009-9450-4

\bibitem{Mikhailov86}
Mikhailov AV 1986 Integrable magnetic models Soliton (Modern Problems in Condensed Matter vol 17) ed S E Trullinger et al (Amsterdam: North-Holland) pp 623-90

\bibitem{MSS}
 Mikhailov A V, Shabat A B and Sokolov V V 1991 The Symmetry Approach to Classification of Integrable Equations, in Nonlinear Dynamics, Ed Zakharov V E, Springer, pp 115-184.
 
\bibitem{Olver93}
Olver P~J 1993 Applications of Lie Groups to Differential Equations, 2nd edn. (New York, Springer). DOI: 10.1007/978-1-4612-4350-2

\bibitem{RS00}
Rogers C and Schief WK 2000 Backlund and Darboux Transformations. Geometry and Modern Applications in Soliton Theory (Cambridge: Cambridge University Press). DOI: 10.1017/CBO9780511606359

\bibitem{s01}
Sakovich S Yu, True and fake Lax pairs: how to distinguish them,  arXiv:nlin.SI/0112027.

\bibitem{s02}
Sakovich S Yu, Cyclic bases of zero-curvature representations: five illustrations to one concept, arXiv:nlin/0212019v1.

\bibitem{Sym82}
Sym A, 1982 Soliton surfaces. Lett. Nuovo Cimento \textbf{33}, 394-400, which also mentions J. Tafel's contribution.
  
\bibitem{Sym85}
Sym A, 1995 Soliton surfaces and their applications (Soliton geometry from spectral problems) Geometric Aspect of the Einstein Equation and Integrable Systems (Lectures Notes in Physics vol 239) ed R Martini (Berlin: Springer) pp 154-231. DOI: 10.1007/3-540-16039-6\_6

\bibitem{Tafel95}
Tafel J 1995 Surfaces in $\mathbb{R}^3$ with prescribed curvature J. Geom. Phys. 17 381-90. \\DOI: 10.1016/0393-0440(94)00054-9

\bibitem{ZM79}
Zakharov V E and Mikhailov A V 1979 Relativistically invariant two-dimensional models of field theory which are integrable by means of the inverse scattering problem method Sov. Phys. -- JETP \textbf{47} 1017--49
 
\bibitem{Zakrzewski89}
Zakrzewski W 1989 Low Dimensional Sigma Models (Bristol: Hilger).

\end{thebibliography}

\end{document}